\documentclass[%
 reprint,
superscriptaddress,
nofootinbib,
 amsmath,amssymb,
 aps,
]{revtex4-2}

\usepackage{graphicx}% Include figure files
\usepackage{dcolumn}% Align table columns on decimal point
\usepackage{bm}% bold math
\usepackage{amsmath}
\usepackage{hyperref}% add hypertext capabilities
\usepackage{tikz}
\usetikzlibrary{arrows.meta}
\usetikzlibrary{decorations.pathmorphing,decorations.markings}
\usepackage{xcolor}

\usepackage{fancyhdr}
\renewcommand{\headrulewidth}{0pt} % 如果不需要頁首那條橫線，請設為 0pt

\begin{document}

\preprint{APS/123-QED}

\title{Alternating Wentzel--Kramers--Brillouin Approximation to the Schr\"{o}dinger Equation: the Bremmer series and beyond}
% Force line breaks with \\

\author{Yu-An Tsai}

 \email{yuantsai@narlabs.org.tw}
 \affiliation{National Center for Instrumentation Research, National Institutes of Applied Research, Hsinchu 300, Taiwan, R.O.C.}
 \affiliation{Institute of Applied Mechanics, National Taiwan University, Taipei 106, Taiwan, R.O.C.}
 \affiliation{Center for Quantum Science and Engineering, National Taiwan University, Taipei 106, Taiwan, R.O.C.}
 %\affiliation{National Taiwan Unversity, Taipei 106, Taiwan, R.O.C.}

 \author{Sheng D. Chao}
 \email{sdchao@iam.ntu.edu.tw}
 \affiliation{Institute of Applied Mechanics, National Taiwan University, Taipei 106, Taiwan, R.O.C.}
 \affiliation{Center for Quantum Science and Engineering, National Taiwan University, Taipei 106, Taiwan, R.O.C.}

\date{\today}% It is always \today, today,

\begin{abstract}
We propose an extension of the Wentzel--Kramers--Brillouin (WKB) approximation for solving the Schr\"{o}dinger equation. %A set of coupled differential equations is obtained by 
Based on an ansatz for the wave function, subject to an auxiliary condition on its first derivative, we obtain a set of 
coupled differential equations that decouple, via an alternating perturbation method, into the well-known Bremmer series. 
The perturbation improves the wave-function amplitudes alternately, and its {\it phase} is refined through recursive diagonalization. From this construction, we derive a general quantization formula that encodes the 
geometric-optics-like physics of the system. When the ratio of the differential reflection coefficient to the classical 
momentum remains constant, the formula reduces to a closed-form quantization condition consistent with that obtained by 
re-summing the WKB series to all orders.
%It is shown that the alternating perturbation method can decouple the set of differential equations, yielding the well-known Bremmer series, 
%and in addition, by virtue of improvements on the amplitudes, it could also refine the {\it phase} of the wave function in 
%a sequence of recursive diagonalizations. 
\end{abstract}

\maketitle
\thispagestyle{plain}
\fancypagestyle{plain}{
  \fancyhf{}
  \fancyhead[R]{\thepage}
  \renewcommand{\headrulewidth}{0pt}
}
%\tableofcontents

\section{\label{sec:Introduction}Introduction}

The Wentzel--Kramers--Brillouin (WKB) theory is one of the most powerful techniques %from which one can 
for obtaining moderately accurate analytical solutions to the Schr\"{o}dinger equation for both eigenvalues and eigenfunctions \cite{Bender_Orszag,DPark}. %It is well known that 
As is well known, the WKB theory works well when the wave function exhibits a rapidly oscillating phase accompanied by a 
slowly varying amplitude; this condition is called the {\it short-wavelength limit} \cite{Sakurai1985}. 
In quantum mechanics, this condition is usually fulfilled in the semiclassical limit ($\hbar \to 0$) \cite{Berry1972}. 
However, when the system is characterized by the {\it long-wave limit} or becomes less {\it semiclassical}, 
the WKB theory tends to be ineffective. In addition, the singular behavior of the wave function surfaces as it approaches 
the classical turning point. Consequently, many studies have addressed, and continue to address, the drawbacks of either 
the wave function or the quantization condition \cite{Friedrich1996phase,Friedrich1996nonintegral, Froman1965jwkb, MillerGood1953, Sueishi}. 
Here we propose a different approach that not only extends the validity of the WKB approximation but also echoes the 
recently published energy-quantization condition derived from the all-orders re-summation of the WKB series
\cite{Tripathi2022}.   

Let us first review the WKB approximation. Consider the time-independent Schrödinger equation in one dimension (the x coordinate)
\begin{equation}\label{eq:1}
\psi''(x) + \frac{p^{2}}{\hbar^{2}}\psi(x) = 0. 
\end{equation} 
with the classical momentum, $p(x) = \sqrt{2m[E-V(x)]}$, where $m$ is the particle mass, and $E$ and $V(x)$ denote the total energy and potential energy, respectively. The first-order WKB approximation gives fundamental solutions
\begin{equation}\label{eq:2}
\psi(x) = \frac{1}{\sqrt{p}}e^{ \pm \frac{i}{\hbar}\int^{x}{pdx'} }.
\end{equation}
Because the Schrödinger equation is a second-order linear differential equation, the complete approximation, with an error of $\mathcal{O}(\hbar)$ relative to the exact solution, should be a linear combination of the two component waves
\begin{equation}\label{eq:3}
\psi(x) = \frac{A}{\sqrt{p}}e^{ \frac{i}{\hbar}\int^{x}{pdx'} } + \frac{B}{\sqrt{p}}e^{ -\frac{i}{\hbar}\int^{x}{pdx'} },
\end{equation}
%%%%%%%%%
where $A$ and $B$ are undetermined constants and complex conjugates of each other. Provided that the phase is integrable 
over the domain of interest, the WKB approximation retains its simple structure regardless of the complexity of the 
potential $V(x)$. Usually, extending the WKB theory becomes difficult as one proceeds to higher-order 
approximations \cite{Bender_Orszag,Kemble1935,Kemble1937}. For example, deriving the connection formulas becomes 
increasingly tedious as the order increases, and the inherent singularity at the turning points, where the classical 
momentum vanishes, leads to a stronger divergence when higher-order terms are naively added to the WKB series. 
Indeed, avoiding mathematical complications beyond the leading order of the WKB approximation is always 
challenging \cite{Beckel1963}; techniques such as Borel summation of exact WKB analysis and the uniform WKB method have 
therefore been developed \cite{Sueishi,Voros1983,Dunne2014}. Recently, Tripathi developed an elegant method that 
re-summed the perturbative WKB series and provided a closed-form quantization formula for two-turning-point problems in 
one dimension \cite{Tripathi2022}. Tripathi's method preserves the simple form of the solution yet greatly improves the 
accuracy of the Bohr-Sommerfeld-Wilson condition, relative to the conventional WKB approximation, over a wider range of 
applicability. This method also reveals the dispersion relation exhibited by the WKB series in terms of geometrical optics. 
Therefore, exploring in detail the geometrical-optics character of the reflected and transmitted 
waves encoded in the WKB series is of significant academic interest.

We propose an extended WKB approximation, dubbed the alternating WKB (a-WKB) method, based on a heuristic ansatz for the wave function in the two-turning-point problem. Sec. \ref{sec:Theory} presents the theoretical approach of the a-WKB method, in which the wave function is corrected as
\begin{equation}\label{eq:4}
\psi(x) = \frac{1}{\sqrt{p}}(1+\int^{x}{\frac{p'}{2p}}e^{-2iS/\hbar }dx')e^{iS/\hbar} + c.c.,
\end{equation}
where c.c. denotes the complex conjugate, $S(x)$ denotes the phase integral, $\int{p}dx$, and $\frac{p'}{2p}$ represents the differential reflection coefficient due to a varying potential $V(x)$. Sec. \ref{sec:Applications} investigates the convergence of the higher-order perturbation series in a harmonic oscillator and discusses the problem of reflections above the potential barrier. Sec. \ref{sec:GeometricOptics} then demonstrates how the a-WKB method can be incorporated into a recursively diagonalizing scheme, in which the wave function is ``observed'' in a new basis to improve the {\it phase}.

Based on the diagonalized wave function obtained in Sec. \ref{sec:GeometricOptics}, Sec. \ref{sec:Quantization} derives the general quantization formula
\begin{equation}\label{eq:5}
\frac{1}{\hbar}\oint_{\Gamma}{ \sqrt{ p^{2}-\left(\frac{\hbar^{2}p'^{2} }{4p^{2}}\right)-\left(\frac{\hbar^{2}\left|\theta'\right|^{2}}{4}\right) - \dots } }dz
= 2(K+1)\pi,
\end{equation}
\\ where $K = (0,1,2\dots)$; here, the phase integral is corrected by subtracting the squared modulus of the differential reflection coefficient, $(\frac{p'}{2p})^{2}$, together with the remainder terms of the sub-reflected waves, $\hbar^{2}\left|\theta'\right|^{2}/4 $ and $[\dots]$, from the square of the primary wave number ($k^{2} = p^{2}/\hbar^{2}$), where the contour $\Gamma$ encircles the two turning points once, counterclockwise. This construction allows us to explain why the {\it odd-order terms} do not contribute to the quantization formula, a phenomenon familiar to mathematical physicists but not clearly understood from a physical perspective. We further explain why, under appropriate conditions, the WKB theory usually provides an accurate approximation even in systems far from the semiclassical limit; under such conditions, the quantization formula (Eq. \eqref{eq:5}) also reduces to the much simpler form
\begin{equation}\label{eq:6}
  \oint_{\Gamma}{ \sqrt{ p^{2}-\left(\frac{\hbar^{2}p'^{2} }{4p^{2}}\right)}}dz = 2(K+1)\pi \hbar, \quad \text{$(K=1,2,\dots).$}
\end{equation}
We also derive the corresponding wave function as the $0$-th order of our diagonalized approximation.
 
Sec. \ref{sec:Conclusion} summarizes our main results, proposes directions for future research, and compares our approach with the theories of Poirier and Floyd, and of Faraggi and Matone, which resemble the a-WKB approximation in form but cover different scopes
\cite{Floyd1982bohr,Floyd1982modified,Faraggi1998equivalence,Poirier2004reconciling}.

%%%%%%%%%%%%%%%

\section{\label{sec:Theory} Theory of The Alternating WKB Approximation}

We first consider a more general ansatz for the wave function than that of Eq. \eqref{eq:3}
\begin{equation}\label{eq:7}
\psi(x) = a(x)e^{iS/\hbar} + b(x)e^{-iS/\hbar},
\end{equation}
where $a(x)$ and $b(x)$ denote the two complex-conjugate amplitude variables, which are not constants. The indefinite phase integral is defined as follows:

\begin{equation}\label{eq:8}
S(x) \equiv \int^{x}{pdx'}.
\end{equation}

To visualize this ansatz, one may treat the right-going and left-going waves, $a  e^{iS/\hbar}$ and $b  e^{-iS/\hbar}$, respectively, as linearly independent solutions that together form a coordinate system describing the wave function in a two-dimensional space. Substituting Eq. (\ref{eq:7}) into Eq. (\ref{eq:1}) under the conditions
\begin{subequations}\label{eq:9}
\begin{align}
&\psi'(x) = i\frac{p}{\hbar}[ a(x)e^{iS/\hbar} - b(x)e^{-iS/\hbar}], \label{eq:9a} \\ 
&\psi''(x) = -\frac{p^{2}}{\hbar^{2}}[ a(x)e^{iS/\hbar} + b(x)e^{-iS/\hbar} ]. \label{eq:9b}
\end{align}
\end{subequations}
This procedure resembles the method of variation of parameters. The first condition implies that the amplitude variations 
$a(x)$ and $b(x)$ of the two interacting waves are sufficiently smaller than their respective phase terms 
(i.e., $a'e^{iS/\hbar}+b'e^{-iS/\hbar}=0$), while the second condition suggests a plane-wave-like solution in which the 
wave number equals the classical momentum, namely, the Schr\"{o}dinger equation itself. Together, these conditions yield 
a set of coupled differential equations for the amplitude variables \footnote{For the derivation of the coupled differential 
equations, see Appendix \ref{Appendix.A}. }
\begin{equation}\label{eq:10}
\left\{
\begin{array}{rcl}
&a'(x) &= -\frac{p'}{2p}a(x) + \frac{p'}{2p}b(x) e^{-2iS/\hbar}, \\ \\
&b'(x) &= -\frac{p'}{2p}b(x) + \frac{p'}{2p}a(x) e^{2iS/\hbar}. 
\end{array} \right.
\end{equation}
The coupled differential equations are interpreted as follows: if a particle moves forward, from left to right along the 
x-axis, the function $a(x)$ describes the forward-going wave amplitude; when the particle encounters a finite potential, 
a reflected wave is generated as it proceeds into the interaction region, and this backward-directed wave amplitude is 
described by the function $b(x)$. The physics of the incoming and reflected waves is governed by the coupling terms, 
in which $\frac{p'}{2p}$ is termed the differential reflection coefficient \cite{Berry1972}. One way to decouple the 
equations is to apply the method of averaging: if the phases are highly oscillating, the contributions of the inhomogeneous 
terms with the factors $e^{2iS/\hbar}$ and $e^{-2iS/\hbar}$ fade into the background. Although this process yields the 
first-order WKB solutions, it entirely ignores the interference effects. To explore the mathematical structure of these 
equations more fully, we instead adopt the perturbation method, first rewriting the coupled differential equations in 
matrix form and introducing a sufficiently small bookkeeping parameter $\varepsilon$:
\begin{equation}\label{eq:11}
  \frac{d}{dx}
  \begin{bmatrix}
  a  \\
  b 
  \end{bmatrix}
  = -\frac{p'}{2p}
  \begin{bmatrix}
  1                 &     -\varepsilon  e^{-2iS/\hbar} \\
  -\varepsilon e^{2iS/\hbar} &      1
  \end{bmatrix}
  \begin{bmatrix}
  a \\
  b
  \end{bmatrix}.
\end{equation}
Considering an expansion of the amplitudes, $a(x)$ and $b(x)$, as a perturbation series in $\varepsilon$:
\begin{equation}\label{eq:12}
  \frac{d}{dx}
  \begin{bmatrix}
  \sum\limits_{n=0}^{\infty}\varepsilon^{n}a_{n}  \\ \\
  \sum\limits_{n=0}^{\infty}\varepsilon^{n}b_{n} 
  \end{bmatrix}
  = -\frac{p'}{2p}
  \begin{bmatrix}
  1                 &     -\varepsilon  e^{-2iS/\hbar} \\
  -\varepsilon e^{2iS/\hbar} &      1
  \end{bmatrix}
  \begin{bmatrix}
    \sum\limits_{n=0}^{\infty}\varepsilon^{n}a_{n}  \\ \\
    \sum\limits_{n=0}^{\infty}\varepsilon^{n}b_{n}
  \end{bmatrix},
\end{equation}
\\ where $a_{n}$ and $b_{n}$ are the perturbed terms. Accordingly, $a_{0}$ and $b_0$ represent the unperturbed (zeroth-order) terms, and the unperturbed equations are written as
\begin{equation}\label{eq:13}
  \frac{d}{dx}
  \begin{bmatrix}
  a_{0} \\
  b_{0}
  \end{bmatrix}
  = -\frac{p'}{2p}
  \begin{bmatrix}
  a_{0} \\
  b_{0}
  \end{bmatrix}.
  \end{equation}
The solutions to Eq. \eqref{eq:13} are readily obtained as
  \begin{equation}\label{eq:14}
  \begin{bmatrix}
  a_{0} \\
  b_{0}
  \end{bmatrix}
  \sim
  \begin{bmatrix}
  \frac{1}{\sqrt{p}} \\ \\
  \frac{1}{\sqrt{p}}
  \end{bmatrix}.
  \end{equation}
These are the amplitudes corresponding to the usual first-order WKB solutions. The first-order perturbation equations ($O[\varepsilon^{1}]$) are expressed as
  \begin{equation}\label{eq:15}
    \frac{d}{dx}
    \begin{bmatrix}
    a_{1} \\
    b_{1}
    \end{bmatrix}
    = -\frac{p'}{2p}
    \begin{bmatrix}
    a_{1} -  b_{0}e^{-2iS/\hbar} \\ \\
    b_{1} -  a_{0}e^{2iS/\hbar}
    \end{bmatrix}.
  \end{equation}
Substituting Eq. \eqref{eq:14} into Eq. \eqref{eq:15}, we obtain
  \begin{equation}\label{eq:16}
    \frac{d}{dx}
    \begin{bmatrix}
    a_{1} \\
    b_{1}
    \end{bmatrix}
    = -\frac{p'}{2p}
    \begin{bmatrix}
    a_{1} - \frac{1}{\sqrt{p}}e^{-2iS/\hbar} \\ \\
    b_{1} - \frac{1}{\sqrt{p}}e^{2iS/\hbar}
    \end{bmatrix}.
    \end{equation}
Eq. \eqref{eq:16} can be solved using the Green's function method in the region $[0, x_{0}]$:
    \begin{equation}\label{eq:17}
    \frac{d}{dx}
    \begin{bmatrix}
    G_{1}(x,\xi) \\ 
    G_{2}(x,\xi)
    \end{bmatrix}
    + \frac{p'}{2p}
    \begin{bmatrix}
    G_{1}(x,\xi) \\ 
    G_{2}(x,\xi)
    \end{bmatrix}
    =
    \begin{bmatrix}
    \delta(x-\xi) \\ 
    \delta(x-\xi)
    \end{bmatrix},
    \end{equation}
where $x_{0}$ denotes the nearest turning point in the neighborhood of the initial point $x = 0$, and the impulse delta function is applied precisely at the point $x = \xi$. Under the homogeneous boundary conditions, we have
    \begin{equation}\label{eq:18}
    \begin{bmatrix}
    G_{1}(0,\xi) \\ 
    G_{2}(0,\xi)
    \end{bmatrix}
    =
    \begin{bmatrix}
    0 \\
    0
    \end{bmatrix}.
    \end{equation}
Applying the homogeneous solutions of Eq. \eqref{eq:17}, we obtain
    \begin{equation}\label{eq:19}
    \begin{bmatrix}
    G_{1}(x,\xi) \\ 
    G_{2}(x,\xi)
    \end{bmatrix}
    =
    \begin{bmatrix}
    \frac{C_{1}}{\sqrt{p}} \\ \\
    \frac{C_{2}}{\sqrt{p}}
    \end{bmatrix}
    \qquad \mbox{for}\qquad 0<x<\xi,
    \end{equation}
    and 
    \begin{equation}\label{eq:20}
    \begin{bmatrix}
    G_{1}(x,\xi) \\ 
    G_{2}(x,\xi)
    \end{bmatrix}
    =
    \begin{bmatrix}
    \frac{C_{3}}{\sqrt{p}} \\ \\
    \frac{C_{4}}{\sqrt{p}}
    \end{bmatrix}
    \qquad \mbox{for}\qquad \xi<x<x_{0},
    \end{equation}
where $C_{1}$ to $C_{4}$ denote undetermined coefficients. Under the homogeneous condition of Eq. \eqref{eq:18}, we obtain $C_{1} = C_{2} = 0$. Applying jump conditions such that
    \begin{equation}\label{eq:21}
      \begin{bmatrix}
      \frac{C_{3}}{\sqrt{p(\xi)}} \\ \\
      \frac{C_{4}}{\sqrt{p(\xi)}}
      \end{bmatrix}
      =
      \begin{bmatrix}
      1 \\
      1
      \end{bmatrix}
      \qquad \mbox{for}\qquad \xi<x<x_{0},
      \end{equation}
we obtain the constants $C_{3}=C_{4}=\sqrt{p(\xi)}$. Consequently, the Green's functions associated with Eq. \eqref{eq:16} are given by

      \begin{equation}\label{eq:22}
      \begin{bmatrix}
      G_{1}(x,\xi) \\ 
      G_{2}(x,\xi)
      \end{bmatrix}
      =
      \begin{bmatrix}
      0 \\
      0
      \end{bmatrix}
      \qquad \mbox{for}\qquad 0<x<\xi,
      \end{equation}
      and 
      \begin{equation}\label{eq:23}
      \begin{bmatrix}
      G_{1}(x,\xi) \\ 
      G_{2}(x,\xi)
      \end{bmatrix}
      =
      \begin{bmatrix}
      \frac{\sqrt{p(\xi)}}{\sqrt{p(x)}} \\ \\
      \frac{\sqrt{p(\xi)}}{\sqrt{p(x)}}
      \end{bmatrix}
      \qquad \mbox{for}\qquad \xi<x<x_{0}.
      \end{equation}
Therefore, the functions $a_{1}(x)$ and $b_{1}(x)$ can be constructed as
      \begin{eqnarray}\label{eq:24}
      \begin{bmatrix}
      a_{1}(x) \\ \\
      b_{1}(x) 
      \end{bmatrix}
      &=
      &\begin{bmatrix}
      \int_{0}^{x_{0}}{G_{1}(x,\xi)\frac{ 1 }{\sqrt{p(\xi)}}\frac{p'(\xi)}{2p(\xi)}e^{-2iS/\hbar}}d\xi \\\\
      \int_{0}^{x_{0}}{G_{2}(x,\xi)\frac{1}{\sqrt{p(\xi)}}\frac{p'(\xi)}{2p(\xi)}e^{2iS/\hbar}}d\xi
      \end{bmatrix}
      \nonumber\\ \nonumber\\
      &=
      &\begin{bmatrix}
      \frac{1}{\sqrt{p(x)}}\int_{0}^{x}{ \frac{p'(\xi)}{2p(\xi)}e^{-2iS(\xi)/\hbar} }d\xi \\\\
      \frac{1}{\sqrt{p(x)}}\int_{0}^{x}{ \frac{p'(\xi)}{2p(\xi)}e^{2iS(\xi)/\hbar} }d\xi 
      \end{bmatrix}.
      \end{eqnarray}
Setting $\varepsilon = 1$, we obtain the first-order perturbations of the amplitudes.
      \begin{eqnarray}\label{eq:25}
      \begin{bmatrix}
      a \\
      b
      \end{bmatrix}
      &=
      &\begin{bmatrix}
      a_{0} + a_{1} \\
      b_{0} + b_{1}
      \end{bmatrix}
      \nonumber\\ \nonumber\\
      &=
      &\begin{bmatrix}
      \frac{ D_{1} }{\sqrt{p}} + \frac{1}{\sqrt{p(x)}}\int_{0}^{x}{ \frac{p'(\xi)}{2p(\xi)}e^{-2iS(\xi)/\hbar} }d\xi    \\ \\
      \frac{ D_{2} }{\sqrt{p}} + \frac{1}{\sqrt{p(x)}}\int_{0}^{x}{ \frac{p'(\xi)}{2p(\xi)}e^{2iS(\xi)/\hbar} }d\xi
      \end{bmatrix}.
      \end{eqnarray}
      \begin{figure}
        \includegraphics[scale=0.95]{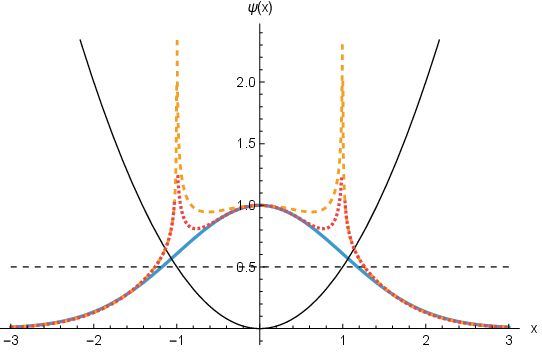}% Here is how to import EPS art
\caption{\label{aWKB_on_HO} Comparison of the WKB (orange dashed line) and a-WKB (red dotted line) approximations for a ground-state harmonic oscillator in atomic units, $m=\omega= \hbar= 1, E_{0}=0.5$ (black dashed line), with the exact solution (blue solid line).}
      \end{figure}
In perturbation theory, the constants $D_{1}$ and $D_{2}$ can reasonably be placed with the unperturbed terms rather than in front of the amplitude functions $a(x)$ and $b(x)$. We therefore choose the constants $D_{1}$ and $D_{2}$ to be the factors $e^{\mp i\frac{\pi}{4}},$\footnote{The factors $e^{\mp i\frac{\pi}{4}}$ will appear more naturally in the derivation in Secs. \ref{sec:GeometricOptics} and \ref{sec:Quantization}.} thereby obtaining the first-order perturbation solutions of the a-WKB theory with the normalization constant $D$:
      \begin{equation}\label{eq:26}
      \begin{aligned}
      \psi_{\text{aWKB} }(x)  
              &= D  \left(\frac{ e^{-i\frac{\pi}{4}} }{\sqrt{p(x)}} + \frac{ 1 }{\sqrt{p(x)}}\int_{0}^{x}{ \frac{p'}{2p}e^{-2iS/\hbar} }
              d\xi\right)e^{iS/\hbar } \\
              &+c.c.
      \end{aligned}
      \end{equation}
\\ The extra correction terms in the amplitudes arise from the reflected sub-waves, which exert a {\it perturbative} effect on the amplitude $1/\sqrt{p(x)}$.
      %

%%%%%%%%%%%%%%%%%%%%%%%%%%%%%%%%%%%%%%%%%%%%%%%%%%%%%%%%%%%%

      \section{\label{sec:Applications} Applications}
      
The improvement in the wave function afforded by the a-WKB method is demonstrated here for the harmonic oscillator problem. 
Because the solutions of the WKB method typically deteriorate at lower energy states of a physical system, we focus on the 
ground-state wave function, as shown in Fig. \ref{aWKB_on_HO}. Without loss of generality, we choose $m =\hbar=\omega = 1$, $E_{0} = 1/2$ 
\footnote{By the method of stationary phase approximation, the correction terms in Eq. \eqref{eq:26} to the amplitude are 
of order $\mathcal{O}(\hbar)$. The approximation with an error of $\mathcal{O}(\hbar)$ gives no improvement over the 
conventional quantization formula of the first order of WKB. Higher-order corrections to the {\it phase} part will be 
demonstrated in Sec. \ref{sec:Quantization}.}, and the classical turning points $x_{0} = \pm 1$, where $m$, $\omega$, and $E_{0}$ represent the mass, angular frequency, and ground-state energy, respectively. 
The amplitudes are clearly seen to improve in a perturbative fashion, and proceeding from the first-order perturbation to 
higher-order terms yields the well-known Bremmer series \cite{Bremmer1951}, which has been studied extensively and shown 
to be a powerful technique for calculating reflection coefficients in potential barrier problems \cite{Berry1972,Berry1982,Maitra1996semiclassical}.
      
\begin{figure}
        \includegraphics[scale=0.53]{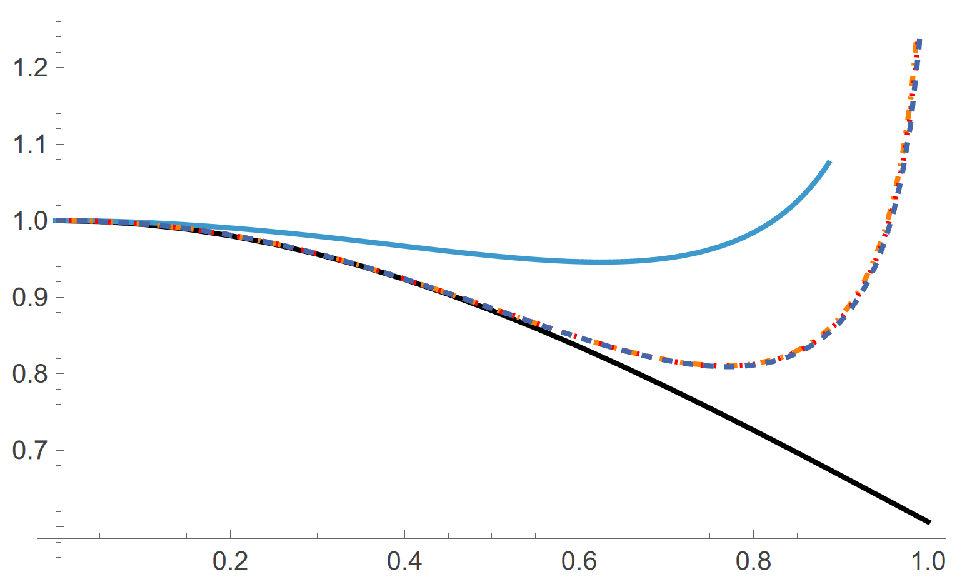}
\caption{\label{3rd_order_aWKB_HO}{A closer examination of the high-order terms of the a-WKB approximation, which modify the wave function of a harmonic oscillator. The blue solid, orange dashed, red dashed, and blue dashed lines represent the first-order WKB, first-order a-WKB, second-order a-WKB, and third-order a-WKB approximations to the wave function, respectively. The black solid line plots the exact solution.}}
      \end{figure}

\begin{figure}
        \includegraphics[scale=0.53]{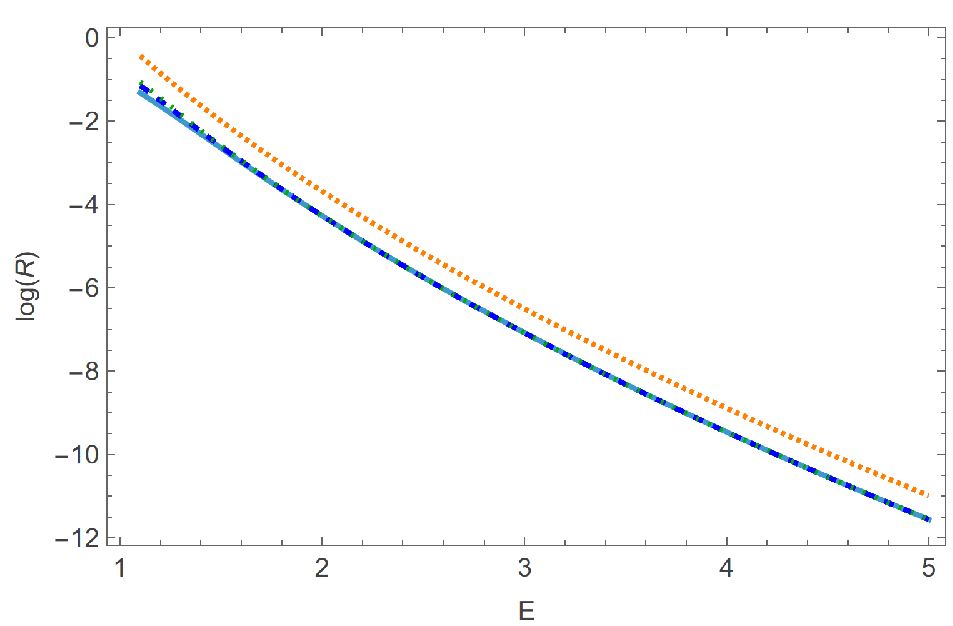}
\caption{\label{3rd_order_aWKB_Sech}{Numerically evaluated reflection coefficients of the potential barrier $\text{Sech}^{2}(x)$, plotted as a function of energy. The reflection coefficients are expressed in logarithmic form to prevent the values from running off the scale. The orange dashed, blue dashed, and green dotted lines plot the first-order WKB, first-order a-WKB, and third-order a-WKB solutions, respectively, while the blue solid line represents the exact solution.}}
      \end{figure}

Here we present the higher-order terms for the bound state of a harmonic oscillator and the scattering state of the $\text{Sech}^{2}(x)$ barrier, as shown in Fig. \ref{3rd_order_aWKB_HO} and Fig. \ref{3rd_order_aWKB_Sech}, respectively. The general high-order terms of the a-WKB wave function are readily obtained as

     \begin{equation} \label{eq:27}
      \begin{aligned}
      &\psi_{\text{aWKB}}(x) = \frac{D}{\sqrt{p}}\bigg[e^{-i\frac{\pi}{4}} + \int_{0}^{x}{dx'\frac{p'}{2p}e^{-2iS/\hbar}}  \\\\
      &+ \int_{0}^{x}{dx'\frac{p'}{2p}e^{-2iS/\hbar}}\int_{0}^{x'}{dx_{1}'\frac{p'}{2p}e^{-2iS/\hbar}}\\ \\
      &+ \int_{0}^{x}{dx'\frac{p'}{2p}e^{-2iS/\hbar}}\int_{0}^{x'}{dx_{1}'\frac{p'}{2p}e^{-2iS/\hbar}}\int_{0}^{x_{1}'}{dx_{2}'\frac{p'}{2p}e^{-2iS/\hbar}}\\\\
      & +\dots  \bigg] e^{iS/\hbar} +c.c.,
      \end{aligned} 
     \end{equation}
\\ Fig. \ref{3rd_order_aWKB_HO} compares the first- to third-order a-WKB wave functions with the first-order WKB wave function and the exact solution for a ground-state harmonic oscillator. Although the higher-order terms of the a-WKB wave functions were found to remain divergent as they approach the turning point ($x_{0}=1$), the second- and third-order a-WKB terms diverge no further than the first-order term. This tendency suggests that the divergence does not arise from the series expansion itself but from the singular behavior of individual terms at the zeros of the momentum or wave number.

The same issue need not concern us when treating scattering problems. According to Eq. (\ref{eq:7}), the reflection coefficient, $R$, should be defined as $|b(-\infty)|^{2}$, which accounts for the backward-going reflected waves; the odd-order terms in $b(x)$ are therefore the relevant candidates in the calculation \cite{Berry1982}. Under the initial conditions $a(-\infty) = 1$ and $b(\infty)=0$, the reflection coefficient $R$ is calculated as
    
    \begin{equation}\label{eq:28}
      \begin{aligned}
    &R = |b(-\infty)|^{2} = \lvert \int_{-\infty}^{\infty}{ \frac{p'}{2p}e^{2iS/\hbar}dx }\\\\
    &+ \int_{-\infty}^{\infty}{ \frac{p'}{2p}e^{2iS/\hbar}dx }\int_{x}^{-\infty}\frac{p'}{2p}e^{-2iS/\hbar}dx_{1}\int_{x}^{\infty}\frac{p'}{2p}e^{2iS/\hbar}dx_{2}\\\\
    &+\dots    \rvert ^{2}.
      \end{aligned}  
  \end{equation}
Applying Eq. (\ref{eq:28}) to the scattering potential, $V(x)=\text{Sech}^{2}(x)$, we evaluated the results numerically, as shown in Fig. \ref{3rd_order_aWKB_Sech}. The convergence of a-WKB proved to be twofold: the first-order a-WKB already provides an extremely accurate approximation to the reflection coefficient as energy increases, consistent with expectations from semiclassical theory, and the first-, second-, and third-order a-WKB approximation curves are almost indistinguishable. This rapid convergence of the Bremmer series to the exact solution is consistent with the behavior deduced for the higher-order a-WKB series in the harmonic oscillator case. \\\\

%%%%%%%%%%%%%%%%%%%%%%%%%%%%%%%%%%%%%%%%%

\section{\label{sec:GeometricOptics} Diagonalization and its Geometrical-Optics Interpretation}

Note that in Eq. \eqref{eq:26}, the stationary phase approximation of the extra perturbation term is of order $\mathcal{O}(\hbar)$, implying that the extra perturbative corrections appear in the amplitude only.
      
What about the phase? This question leads us to the following three questions:
      
\begin{enumerate}
\item [(1)] Does the a-WKB approximation to high order of $\hbar$ also improve the phase part of the wave function? \item [(2)] Why do the {\it odd-order terms} in the WKB series not contribute to the quantization formula? \item [(3)] Why does the WKB approximation accurately predict the energy eigenvalue in many cases?
\end{enumerate}

Given that the coupled differential equations, i.e., Eq. \eqref{eq:10}, can reform the wave function by adding perturbative sub-waves, we take the first derivative of Eq. \eqref{eq:7} and extend the validity of the constraint in Eq. \eqref{eq:9a}, including the amplitude variations $a(x)$ and $b(x)$:
\begin{equation}\label{eq:29}
  \begin{aligned}
\psi'(x) = [a'(x)+ \frac{i}{\hbar}a \cdot p(x)]e^{iS/\hbar} + [b'(x)- \frac{i}{\hbar}b \cdot p(x)]e^{-iS/\hbar}.\\
  \end{aligned}
\end{equation}
\\ Iterating the coupled differential equations, Eq. \eqref{eq:10}, into Eq. \eqref{eq:29}, we obtain
\begin{equation}\label{eq:30}
  \begin{aligned}
    \psi'(x) = &\left[\frac{-p'}{2p}a+\frac{p'}{2p}be^{-2iS/\hbar} + \frac{i}{\hbar}a \cdot p(x)\right]e^{iS/\hbar} \\
             + &\left[\frac{-p'}{2p}b+\frac{p'}{2p}ae^{-2iS/\hbar}- \frac{i}{\hbar}b \cdot p(x)\right]e^{-iS/\hbar}.
  \end{aligned}
\end{equation}
Under the conditions of Eq. \eqref{eq:9a}, Eq. \eqref{eq:30} does not differ from the coupled differential equations, Eq. \eqref{eq:10}. However, as suggested by Eq. \eqref{eq:7}, it is instructive to define two variables, $\psi_{+}$ and $\psi_{-}$, referring respectively to waves traveling to the left and to the right, as a coordinate system; we analyze them separately in matrix form to relax the constraint
\begin{equation}\label{eq:31}
  \frac{d\vec{\psi}}{dx}
  \equiv
 \frac{d}{dx}
  \begin{bmatrix}
    \psi_{+} \\
    \psi_{-} \\
  \end{bmatrix}
    =
  \begin{bmatrix}
     -\frac{p'}{2p} + \frac{i}{\hbar}p & \frac{p'}{2p} \\
     \frac{p'}{2p}                     & -\frac{p'}{2p} + \frac{i}{\hbar}p
  \end{bmatrix}
  \begin{bmatrix}
    \psi_{+} \\
    \psi_{-} \\
  \end{bmatrix}.
\end{equation}
If only the diagonal terms governing the phase contribution are retained, the first-order WKB approximation is recovered; 
our present analysis instead accounts for the off-diagonal terms. Solving Eq. \eqref{eq:31} is equivalent to identifying 
a suitable coordinate system by means of diagonalization. The $\vec{\psi}$ can be interpreted geometrically as a two-level 
system related to the {\it Bloch sphere} and can be described mathematically as a {\it spinor} \cite{Kampen,Feynman1957}. 
To extract further information from the off-diagonal terms, we adopt the methodology introduced by 
Van Kampen \cite{Kampen}, in which the amplitude information is absorbed through the following transformation:

\begin{equation}\label{eq:32}
\vec{\psi}(x) = \frac{1}{\sqrt{p}}\vec{v}(x) .
\end{equation}
Eq. \eqref{eq:31} then becomes
\begin{equation}\label{eq:33}
  \frac{d\vec{v}}{dx} = 
  \begin{bmatrix}
    \frac{i}{\hbar}p  &  \frac{p'}{2p} \\\\
    \frac{p'}{2p}     &  -\frac{i}{\hbar}p
  \end{bmatrix}
  \vec{v}.
\end{equation}
Using the Pauli matrices, Eq. \eqref{eq:33} is rewritten as
\begin{equation}\label{eq:34}
  \begin{aligned}
    &\frac{d\vec{v}}{dx} = (\frac{ip}{\hbar}\sigma_{z} + \frac{p'}{2p}\sigma_{x}) \vec{v} \\
    &= \frac{i}{\hbar}\sqrt{ p^{2}-\frac{\hbar^{2}p'^{2}}{4p^{2}} } 
    \left( \frac{ \frac{ip}{\hbar} }{\frac{i}{\hbar}\sqrt{ p^{2}-\frac{\hbar^{2}p'^{2}}{4p^{2}} }}\sigma_{z} 
    +  \frac{ \frac{p'}{2p} }{\frac{i}{\hbar}\sqrt{ p^{2}-\frac{\hbar^{2}p'^{2}}{4p^{2}} } } \sigma_{x} \right) \vec{v} \\
  &\equiv 
  \frac{i\tilde{p}}{\hbar} ( \hat{n}\cdot\vec{\sigma} )\vec{v}.
  \end{aligned}
\end{equation}
\\ The effective momentum (or effective wave number)\footnote{See how the differential reflection coefficient $\frac{p'}{2p}$ is incorporated into the {\it phase} part in Eq. \eqref{eq:32}. Analogously to geometrical optics, this indicates that successful diagonalization retains the transmitted part of the traveling wave.}, the unit vector $\hat{n}$, and the vector operator $\vec{\sigma}$ are respectively defined as
\begin{equation}
  \begin{aligned}
  &\tilde{p}(x) \equiv \sqrt{p^{2}- \frac{\hbar^2p'^{2}}{4p^{2}}}, \\
  &\hat{n} \equiv \frac{ \frac{ip}{\hbar} }{ \frac{i}{\hbar}\sqrt{ p^{2}-\frac{\hbar^{2}p'^{2}}{4p^{2}} } }\vec{e_{z}} 
+ \frac{ \frac{p'}{2p} }{ \frac{i}{\hbar}\sqrt{ p^{2}-\frac{\hbar^{2}p'^{2}}{4p^{2}} } }\vec{e_{x}}, \\\\
  &\vec{\sigma} = \sigma_{x} \vec{e_{x}} + \sigma_{y} \vec{e_{y}} + \sigma_{z} \vec{e_{z}},
  \end{aligned} \nonumber
\end{equation}
where the Pauli matrices are given by
\begin{eqnarray}
  \sigma_{x} =
  \begin{bmatrix}
    0  &  1 \\
    1  &  0
  \end{bmatrix} \nonumber
  , \quad
  \sigma_{y} =
  \begin{bmatrix}
    0  &  -i \\
    i  &   0
  \end{bmatrix}
  , \quad
  \sigma_{z} = 
  \begin{bmatrix}
    1 & 0 \\
    0 & -1
  \end{bmatrix}.
\end{eqnarray}
We adopt a geometrical picture in which the unit vector $\hat{n}$ lies in the $\sigma_{x}$-$\sigma_{z}$ plane, as shown in Fig. {\ref{pauli}}, with $( \frac{ip}{\hbar} )/ (\frac{i \tilde{p}}{\hbar})$ and $( \frac{p'}{2p})/ (\frac{i \tilde{p}}{\hbar}) $ denoting the components along the $\sigma_{z}$ and $\sigma_{x}$ axes, respectively, and with the angle $\theta(x)$ defined as
\begin{equation}\label{eq:35}
  \tan\theta \equiv -\frac{i\hbar p'}{2p^{2}}.
\end{equation}
\definecolor{internationalkleinblue}{rgb}{0.0, 0.18, 0.65}
\begin{figure}
  \begin{tikzpicture}[>=Latex,scale=0.9]
   %\draw[help lines] grid(10,10) ;
   \draw[,line width=1,->,shorten >=0.5mm] (5,5) -- (9,5) node[below right,font=\LARGE]{$\sigma_{y}$};
   \draw[line width=1,->] (5,5) -- ([turn] 180:4) node[below right,font=\LARGE]{$\sigma_{x}$};
   \draw[line width=1,->] (5,5) --(5,11) node[right,font=\LARGE]{$\sigma_{z}$};
   \fill[internationalkleinblue] (5,5) circle(0.1);
     \draw[line width=1,internationalkleinblue,->] (5,5)--([turn]75:4) coordinate (Beta);
     \node[font=\Large,text=internationalkleinblue] at (2.5,8.5){$\hat{n}$};
     \node[font=\Large,text=red!70!black] at (6.5,9){$\hat{n}_{2}$};
     \draw[line width=0.5,dashed,opacity=0.6] (Beta)--++(-90:5.4); %vertical dashed line
     \draw[line width=0.5,dashed,opacity=0.6] (Beta)--++(45:2.8); %paralle dashed line
     \node[font=\LARGE,text=internationalkleinblue] at (1.8,6){$(\frac{ip}{\hbar})/(\frac{i\tilde{p}}{\hbar})$};
     \node[font=\LARGE,text=internationalkleinblue] at (3.0,10){$(\frac{p'}{2p})/(\frac{i\tilde{p}}{\hbar})$};
   \draw[line width=1,red!70!black,opacity=0.3,dashed,->] (5,5)--([turn]60:4);
   \node[font=\Large,text=red!70!black] at (4.5,9){$\hat{n}'_{2}$};
     \begin{scope}
    \path[clip] (5,10)--(5,5)--([turn]210:5);
    \fill[internationalkleinblue,opacity=0.5] (5,5) circle(10mm);
    \node[font=\large,text=internationalkleinblue] at (4.8,6.3) {$\theta$};
   \end{scope}
      \draw[line width=1,red!70!black,->] (5,5)--([turn]30:4);
      \draw[dotted,red] (5,8) circle[x radius=1.5,y radius=0.2];
        \path (5,8) ++(0:1.5) arc[start angle=0,end angle=50,x radius=1.5,y radius=0.2] coordinate (starting);
        \fill[gray] (starting) circle(0pt);
        \draw[black!20!red,opacity=1,->] (starting) arc[start angle=50,end angle=-230,x radius=1.5, y radius=0.2];
   \begin{scope}
        \path[clip] (5,10)--(5,5)--([turn]165:4);
        \fill[red!70!black,opacity=0.5] (5,5) circle(12mm);
        \node[font=\large,text=red!70!black] at (5.25,7) {$\theta_{2}$};
   \end{scope}
\end{tikzpicture}
\caption{\label{pauli} The unit vector $\hat{n}$ lies in the $\sigma_{x}$-$\sigma_{z}$ plane, in a three-dimensional space constructed from the Pauli matrices. The quantities associated with $\hat{n}$ are marked in blue. The unit vector $\hat{n_{2}}$ (red) is obtained after one diagonalization step, in the $\sigma_{y}$-$\sigma_{z} $ plane, where $\theta_{2}$ denotes the angle between $\hat{n}_{2}$ and the $\sigma_{z}$ axis. A rotation of $\frac{\pi}{2}$ about the $\sigma_{z}$ axis yields $\hat{n}'_{2}$.}
\end{figure}
The angle $\theta(x)$ is small when the potential $V(x)$ varies slowly. To diagonalize the matrix $\tilde{n}\cdot\vec{\sigma} $ in Eq. \eqref{eq:34}, we rotate the $\sigma_{z}$ axis counterclockwise through the angle $\theta$, about the $\sigma_{y}$ axis, thereby aligning it with $\hat{n}$:
\begin{equation}\label{eq:36}
  \vec{v} = e^{-\frac{i}{2}\theta\sigma_{y}}\vec{w}.
\end{equation}
Substituting Eq. \eqref{eq:36} into Eq. \eqref{eq:34}, we obtain
\begin{equation}\label{eq:37}
  \begin{aligned}
  \frac{d\vec{w}}{dx} &= \left[ \frac{i\tilde{p}}{\hbar}\cdot e^{\frac{i}{2}\theta \sigma{y} }(\hat{n}\cdot \vec{\sigma})e^{-\frac{i}{2}\theta \sigma_{y}} 
  + \frac{i}{2}\frac{d\theta}{dx}\sigma_{y} \right]\vec{w} \\ \\
  &= \left( \frac{i\tilde{p}}{\hbar}\cdot \sigma_{z} + \frac{i}{2}\frac{d\theta}{dx}\sigma_{y} \right)\vec{w}.
  \end{aligned}
\end{equation}
As the {\it spin axis} associated with $\vec{w}$ (here denoted as $\hat{n}_{2}$) lies in the $\sigma_{y}$-$\sigma_{z}$ plane, we rotate the angle by $\frac{\pi}{2}$ along the $\sigma_{z}$ axis, counterclockwise. Letting
\begin{equation} \label{eq:38}
  \vec{w}(x) = e^{ -i\frac{\pi}{4}\sigma_{z} }\tilde{w}.
\end{equation}
We can recast Eq. \eqref{eq:37} into the original form with the variable vector $\tilde{w}$:
\begin{equation} \label{eq:39}
  \frac{d\tilde{w}}{dx} = \left(\frac{i\tilde{p}}{\hbar}\cdot \sigma_{z} + \frac{i}{2}\frac{d\theta}{dx}\sigma_{x} \right)\tilde{w}.
\end{equation}
To explore the underlying mechanism of this methodology, we proceed to the next diagonalization step:
\begin{equation}\label{eq:40}
  \frac{d\tilde{\chi}}{dx} = \left(\frac{i}{\hbar}\sqrt{ \tilde{p}^{2}-\frac{\hbar^{2}|\theta'|^{2}}{4} } \cdot\sigma_{z} + \frac{i}{2}\frac{d\theta_{2}}{dx}\sigma_{x} \right)\tilde{\chi}.
\end{equation}
Here, $\theta_{2}$ denotes the angle between the {\it spin axis} associated with $\tilde{w}$ (here denoted as $\hat{n}'_{2}$) and the $\sigma_{z}$ axis, and the differential reflection coefficient $\frac{p'}{2p}$ is replaced by $\frac{i}{2}\frac{d\theta_{2}}{dx}$.

Eq. \eqref{eq:39} and Eq. \eqref{eq:40} show that the first and second diagonalization steps mathematically modify the classical momentum $p$ to $ \tilde{p} \equiv \sqrt{p^{2}-\frac{\hbar^{2}}{4p^{2}}} $ and $\sqrt{\tilde{p}^{2}- \frac{\hbar^{2}|\theta'|^{2}}{4}}$, respectively; physically, each diagonalization step yields a ``new'' wave number, the remaining principal part, as the wave propagates. The term in the square root carrying a ``{ \it minus}'' sign represents the reflected wave.

Dropping the off-diagonal terms in Eq. \eqref{eq:39}, the diagonalized approximation to first order in $\theta$ is given by
\begin{equation}\label{eq:41}
  \tilde{w}^{(1)}(x) = 
  \begin{bmatrix}
    e^{\frac{i}{\hbar}\int^{x}{ \sqrt{p^{2}-\frac{\hbar^{2}p'^{2}}{4p^{2}}} }dt} \\\\
    e^{-\frac{i}{\hbar}\int^{x}{\sqrt{p^{2}-\frac{\hbar^{2}p'^{2}}{4p^P{2}}}}dt} 
  \end{bmatrix}.
\end{equation}
Recall that the corresponding approximation after one diagonalization step gives the wave function
\begin{equation}\label{eq:42}
  \vec{\psi} = \frac{1}{\sqrt{p}}\vec{v} = \frac{1}{\sqrt{p}}e^{-\frac{i}{2}\theta\sigma_{y}}e^{-i\frac{\pi}{4}}\tilde{w}.
\end{equation}
Therefore, to first order, the diagonalized approximation to the wave function is
\begin{equation}\label{eq:43}
  \begin{aligned}
  &\vec{\psi}^{(1)}(x) \equiv \frac{1}{\sqrt{p}}\vec{v} = \frac{1}{\sqrt{p}}(e^{-\frac{i}{2}\theta \sigma_{y}} e^{ -i\frac{\pi}{4}\sigma_{z} })\cdot \tilde{w}^{(1)}(x) \\\\
  &=\frac{1}{\sqrt{p}} 
  \begin{bmatrix}
    \cos{(\frac{\theta}{2})}e^{-i\frac{\pi}{4}} &  -\sin{(\frac{\theta}{2})}e^{i\frac{\pi}{4}} \\\\
    \sin{(\frac{\theta}{2})}e^{-i\frac{\pi}{4}} &   \cos{(\frac{\theta}{2})}e^{i\frac{\pi}{4}}
  \end{bmatrix}
  \begin{bmatrix}
  e^{ \frac{i}{\hbar}\int^{x}{ \sqrt{p^{2}-\frac{\hbar^{2}p'^{2}}{4p^{2}}} }dt   } \\\\
  e^{ -\frac{i}{\hbar}\int^{x}{ \sqrt{p^{2}-\frac{\hbar^{2}p'^{2}}{4p^{2}}} }dt   }
  \end{bmatrix} .
\end{aligned}
  \end{equation}\\
Unlike the constants $D_{1}$ and $D_{2}$ in Sec. \ref{sec:Theory}, the transformation matrix $e^{ -i\frac{\pi}{4}\sigma_{z} }$ in the first-order diagonalization naturally incorporates the phase factors $ e^{\mp i\frac{\pi}{4}}$.

Analogous to geometric optics, Eqs. \eqref{eq:34}--\eqref{eq:43} demonstrate the appearance of the reflection effect in the amplitude and phase of the wave function, with additional perturbative modifications introduced at each diagonalization step. This recursive process answers the first question posed at the beginning of this section.

%%%%%%%%%%%%%%%%%%%%%%%%%%%%%%%%%%%%%%%%%%
\section{\label{sec:Quantization} The WKB Quantization Condition}

In Sec. \ref{sec:Applications}, we showed the rapid convergence of the Bremmer series \cite{Atkinson1960,Berry1972}. Moreover, as proven by Keller and Keller for a system of first-order linear differential equations with a general $j$-component column vector, Eq. \eqref{eq:31}, the Bremmer-like series converges absolutely and uniformly over a wide region of the complex plane
\cite{Keller1962}. Motivated by these results, we found that a sequence of diagonalization steps can accurately approximate the {\it{phase}} of the wave function. 
Through the recursively diagonalized scheme (Eqs. \eqref{eq:34}--\eqref{eq:43}), the diagonalized a-WKB wave function is derived to the $n$-th order as
 \begin{equation}\label{eq:44}
  \begin{aligned}
  &\vec{\psi}^{(n)} \equiv\begin{bmatrix}
    \psi^{(n)}_{+} \\\\
    \psi^{(n)}_{-}
  \end{bmatrix} 
  = 
  \frac{1}{\sqrt{p}}\vec{v} \\
  &= \frac{1}{\sqrt{p}}(e^{-\frac{i}{2}\theta \sigma_{y}} e^{ -i\frac{\pi}{4}\sigma_{z} }
  e^{ \frac{i}{2}\theta_{2}\sigma_{y} }\dots) \cdot \tilde{\chi}^{(n)} \\
  &=\frac{1}{\sqrt{ p }}(e^{-\frac{i}{2}\theta \sigma_{y}} e^{ -i\frac{\pi}{4}\sigma_{z} }e^{ \frac{i}{2}\theta_{2}\sigma_{y} }\dots)
  \begin{bmatrix}
  e^{  \frac{i}{\hbar}\int^{x}{ p^{2}_{n} }dt}     \\\\
  e^{  -\frac{i}{\hbar}\int^{x}{  p^{2}_{n} } dt}    \\
  \end{bmatrix}, \\
\end{aligned}  
  \end{equation} 
\\ where $p_{n}$ denotes the $n$-fold modified momentum with an $n$-fold correction from reflection. In this notation, $p_{0} = p$, $\tilde{p} = \sqrt{p^{2}-\frac{\hbar^{2}p'^{2}}{4p^{2}}} \equiv p_{1}$, and higher folds are defined similarly.

In particular, the $1$st-order diagonalized approximations to the left- and right-going wave functions are given by
  \begin{equation} \label{eq:45}
    \begin{aligned}
    &\psi^{(1)}_{\pm}(x) = \frac{1}{\sqrt{p}}( \cos\frac{\theta}{2} \pm \sin\frac{\theta}{2} )e^{ \pm \frac{i}{\hbar}\left( \int^{x}{  \tilde{p}}dt -\frac{\pi}{4}\right)  } \\\\
    &=\frac{1}{\sqrt{p}}( 1-\frac{\hbar^{2}p'^{2}}{4p^{4}} )^{-1/4}\left\{ \left[ \frac{1}{2}( 1-\frac{\hbar^{2}p'^{2}}{4p^{4}} )^{1/2} + \frac{1}{2}  \right]^{1/2} \right.\\\\
    &+ \left.\left[ \frac{1}{2}( 1-\frac{\hbar^{2}p'^{2}}{4p^{4}} )^{1/2} - \frac{1}{2} \right]^{1/2}   \right\}e^{ \pm \frac{i}{\hbar}\left( \int^{x}{  \tilde{p}}dt -\frac{\pi}{4}\right)  } \\\\
    &\sim \frac{1}{\sqrt{\tilde{p}}}e^{ \pm i \left(\frac{1}{\hbar}\int^{x}{ \sqrt{ \tilde{p}^{2}-\frac{\hbar^{2}\left|\theta'\right|^{2}}{4}}}dt -\frac{\pi}{4}\right)}, \text{ as $\left| \frac{\hbar p'}{2p^{2}}  \right| \ll \mathcal{O}(1)$}.
  \end{aligned} 
  \end{equation}
We deduce that higher-order diagonalization yields
\begin{equation}\label{eq:46}
  \psi^{(n)}_{\pm}(x) \sim \frac{1}{\sqrt{p_{n}}}e^{ \pm \frac{i}{\hbar}\int^{x}{  p_{n}dt }}, \text{ as $\left| \frac{\hbar p'}{2p^{2}}  \right| \ll \mathcal{O}(1)$}.
\end{equation}
To evaluate the quantization formula associated with Eq. \eqref{eq:44} or Eq. \eqref{eq:46}, let us consider a wave function of the form
  \begin{equation} \label{eq:47}
    \psi \sim Ae^{\frac{i}{\hbar}\tilde{S}} = e^{ \frac{i}{\hbar}[\tilde{S}+ \frac{\hbar}{i}\ln(A) ] },
  \end{equation}
where $\tilde{S}$ and $A(x)$ denote the phase part and amplitude (which can be multi-valued, primarily owing to the square-root function), respectively. To maintain a single-valued wave function, the phase difference over a contour integral ($\Gamma$) must fulfill the condition formulated by Keller \cite{Keller1958,Stone2005}
\begin{equation}\label{eq:48}
  \begin{aligned}
  &\Delta \tilde{S} + \frac{\hbar}{i}\Delta \ln(A) =2K\pi\hbar, \quad (K=0,1,2,\dots).
  \end{aligned}
\end{equation}
If $\tilde{S}$ can be expanded as a perturbation series in $\hbar$, and $A(x)=1$, we obtain
\begin{equation}\label{eq:49}
  \Delta \sum\limits_{n=0}^{\infty}{ \hbar^{n}\tilde{S}_{n} } = 2K\pi\hbar.
\end{equation}

\begin{figure}
  \begin{tikzpicture}[>=Latex,scale=0.8]
 \tikzset{
  branch point/.style={cross out,draw=black,fill=none,minimum size=2*(#1-\pgflinewidth),inner sep=0pt,outer sep=0pt}, 
  branch point/.default=5
}
\tikzset{
  branch cut/.style={
    decorate,decoration=snake
    }}
  %\draw[help lines] grid (10,10);
  %Axes
  \draw[line width=1] (0,5) -- (10,5) node[below right,font=\large]{Re};
  \draw[line width=1] (5,1.5)--(5,8.5) node[below right,font=\large]{Im};
  \fill[internationalkleinblue,opacity=1] (8,5) circle(1.2mm) ;
  %Turning points
  \node[below right,font=\Large] at (8.2,4.8){$z_{1}$};
  \fill[internationalkleinblue,opacity=1] (2,5) circle(1.2mm) ;
  \node[below right,font=\Large] at (1.3,4.8){$z_{2}$};
  %Branch cut
  \draw[branch cut,line width=1,internationalkleinblue] (2,5)--(8,5);
  %Contour
  \draw[xshift=-5,red!70!black,decoration={markings,mark=between positions 0.125 and 0.875 step 0.15 with \arrow{>}},postaction={decorate}] (1,4) -- (9,4) arc (-90:90:1)--(1,6) arc(90:270:1) ;
  \node[font=\LARGE] at (8,8){z};
  \draw[] (8,8) circle[x radius=0.5, y radius=0.5]; 
  \node[red!70!black,font=\large] at (8,6.5){$\Gamma$};
\end{tikzpicture}

\caption{\label{contour} Potential well problem in the complex z-plane. The turning points $z_{1}$ and $z_{2}$ are joined by a branch cut (blue wavy line). In a single-valued evaluation, the contour $\Gamma$ encircles both turning points.}
\end{figure}
Evaluating once around the contour, the phase difference is given as
\begin{equation}\label{eq:50}
  \frac{1}{\hbar} \oint_{\Gamma}{ \sum\limits_{n=0}^{\infty}{\hbar^{n}\tilde{S}'_{n}(z) }}dz = 2K\pi,
\end{equation}
where z is a complex variable. We thus obtain the closed-form quantization formula first derived by Dunham \cite{Dunham1932} and subsequently analyzed in detail by Bender {\it et al.} \cite{Bender1977numerological,Robnik2000some}. This formula underpins the derivation, in the present paper, of the quantization formula associated with Eq. \eqref{eq:44} or Eq. \eqref{eq:46}. As the first-order WKB gives $\tilde{S}'(z)=p(z)$ and $A(x) = \frac{1}{\sqrt{p}}$, we have
  \begin{equation}\label{eq:51}
    \oint_{\Gamma}{ \left[\frac{p}{\hbar}-\frac{1}{2}\ln(p)\right] }dz = 2K\pi  .
  \end{equation}
For the potential well problem, we evaluate once around the contour that encircles the two turning points (two zeros of $p(x)$), as illustrated in Fig. \ref{contour}. Applying the residue theorem such that
\begin{eqnarray}\label{eq:52}
  \frac{1}{\hbar}\oint_{\Gamma}{ p  }dz + 2\pi i( \frac{1}{2}+\frac{1}{2} )=2K\pi .
\end{eqnarray}
\\ We retrieve the conventional quantization formula of the first-order WKB approximation:
\begin{equation}\label{eq:53}
  \frac{1}{\hbar}\oint_{\Gamma}{ p  }dz  =(2K+1)\pi \qquad (K=0,1,2 \dots).
\end{equation}
For the $n$-th-order diagonalized a-WKB wave function, the phase $\tilde{S}$ and amplitude $A(x)$ are respectively given as
\begin{equation}\label{eq:54}
    \tilde{S}'(z) = p_{n},
\end{equation}

and

\begin{equation}\label{eq:55}
    A(x) = \frac{1}{\sqrt{p}}(e^{-\frac{i}{2}\theta \sigma_{y}} e^{ -i\frac{\pi}{4}\sigma_{z} }e^{ \frac{i}{2}\theta_{2}\sigma_{y} }\dots) .
\end{equation}
Using the results derived above, we derive the $n$-th-order quantization formula of the diagonalized a-WKB wave function \footnote{In Appendix \ref{Appendix.B}, we demonstrate that the total phase contribution from the amplitude is $i\pi$.}
\begin{widetext}
\begin{equation}\label{eq:56}
  \begin{aligned}
    \frac{1}{\hbar}\oint_{\Gamma}{ p_{n}dz } 
    = \frac{1}{\hbar}\oint_{\Gamma}{ \sqrt{ \left(p^{2}-\frac{\hbar^{2}p'^{2}}{4p^{2}} - (\frac{\hbar^{2}\left|\theta'\right|}{2})^{2} - 
    \dots - \frac{\hbar^{2}\left| \theta'_{n-1} \right|^{2}}{4} \right) } }dz 
    =(2K+1)\pi, \qquad (K=0,1,2\dots).
  \end{aligned}
\end{equation}
\end{widetext}
As the angle $\theta$ in Eq. \eqref{eq:35} includes $\hbar$, each correction provides two higher orders of $\hbar$ than the previous correction. Eq. \eqref{eq:56} can then be written as
\begin{equation}\label{eq:57}
  \begin{aligned}
    &\frac{1}{\hbar}\oint_{\Gamma}{ \sqrt{ \left(p^{2}-\frac{\hbar^{2}p'^{2}}{4p^{2}} - \mathcal{O}(\hbar^{4}) - \mathcal{O}(\hbar^{6})-\dots \right) } }dz \\\\
    &=(2K+1)\pi, \qquad\qquad\qquad (K=0,1,2\dots).
  \end{aligned}
\end{equation}
\\ In the geometric-optics analogy, the integrand of Eq. \eqref{eq:56} represents a ``ray'' or ``wavefront'' whose effective wave number equals the transmitted part of the primary wave number ($k(x)=p(x)/\hbar$). The classical momentum is reduced by the differential reflection coefficient and by the subsequent $n$ folds of reflected sub-waves, each of which introduces an additional correction of order $\mathcal{O}(\hbar^{2})$.

All {\it odd-order terms} beyond the first order can be formulated as exact derivatives and are therefore excluded from the quantization condition \cite{Froman1965jwkb,Bender1977numerological}; nevertheless, the physical interpretation of this exclusion has remained unclear. As evidenced by the diagonalized wave function of $\psi^{(n)}_{\pm}$, i.e., Eq. \eqref{eq:44}, and by the quantization formula Eq. \eqref{eq:57}, the wave-number improvement is contributed only by even orders of $\hbar$ in the geometric-optics series. This observation offers insight into the second question posed at the beginning of Sec. \ref{sec:GeometricOptics}.

If $\theta$ remains constant along some path $\Gamma'$, the quantization formula given by Eq. \eqref{eq:57} reduces to the quantization condition derived by Tripathi:
 \begin{equation} \label{eq:58}
  \frac{1}{\hbar}\oint_{\Gamma'}{ \sqrt{ \left(p^{2}-\frac{\hbar^{2}p'^{2}}{4p^{2}} \right) } }dz =(2K+1)\pi
\end{equation}
To obtain this closed-form quantization condition, Tripathi re-summed the WKB series to all orders, which was shown to predict the exact energy spectrum of the three-dimensional harmonic oscillator, the Coulomb potential, the Eckart potential, and the Morse potential.

Furthermore, for a constant C along the path $\Gamma'$, we have
\begin{equation} \label{eq:59}
  \frac{ p'}{2p^{2}} \equiv C_{\text{along the path $\Gamma'$}}.
\end{equation}
Integrating along the path $\Gamma'$, we obtain
\begin{equation}\label{eq:60}
  \int_{\Gamma'}{\frac{p'}{2p^{2}}}dx = \int_{\Gamma'}{C}dx.
\end{equation}
Therefore, we have
\begin{equation}\label{eq:61}
   p(x)\cdot x \bigg{|}_{\text{along the path $\Gamma'$}} = -\frac{1}{2C}.
\end{equation}
As $\theta(x)$ behaves as a constant, the integrand of the reduced quantization formula of Eq. \eqref{eq:58} necessarily satisfies the {\it Heisenberg uncertainty principle} for the local momentum and position variables
\begin{equation} \label{eq:62}
 p \cdot x \ge \frac{\hbar}{2}.
\end{equation}
Accordingly, the constant C is upper-bounded as $|C| \leq 1/\hbar$. For a harmonic oscillator ($p(x) = \sqrt{1-x^{2}}$), the contour can be extended to infinity, and the wave number is corrected to first order as follows:
\begin{equation}\label{eq:63}
  \frac{\hbar^{2}p'^{2}}{4p^{2}} = \frac{\hbar^{2}}{4}\frac{1}{x^{2}}.
\end{equation}
Because this correction approaches zero faster than $1/x$ as $x \to \infty$, even when $\hbar=1$, the integrand of Eq. (\ref{eq:56}) forms an asymptotic series. This result is consistent with that obtained by Bender {\it et al.}, who showed that the WKB theory effectively yields the eigenvalues of the power-law potential, $x^{\text{2N}}$, even when the problem is far from the semiclassical limit, that is, $\hbar$ need not approach zero.
\cite{Bender1977numerological,Robnik2000some, Krieger1967, Rosenzweig1968exact}. In general, when $\theta$ is not constant, we must impose either the semiclassical limit ($\hbar \rightarrow 0$) or the complete 
quantization formula Eq. \eqref{eq:56}. We note that the $n$-th-order correction involves an $n$-tuple of differentiations
\begin{equation}\label{eq:64}
  \theta'_{n} \sim \left(\frac{\hbar\cdot\theta'_{n-1}}{2p_{n-1}}\right)' = \hbar\left[ \frac{\theta'_{n-1}}{2p^{2}_{n-1}}+\frac{\theta''_{n-1}}{2p_{n-1}} \right], \text{as $\theta \ll \mathcal{O}(1)$}.
\end{equation}
\\ When the classical momentum $p(x) = \sqrt{E-V(x)}$ has finite derivatives, one can find a specific order of $N$ (and hence any order larger than $N$) such that $\theta'_{N}$ is much smaller than $\mathcal{O}(1)$, yielding an arbitrarily accurate approximation to the eigenenergy of the problem. In particular, the {\it short-wavelength limit} is a special case in which
\begin{equation}\label{eq:65}
  \frac{\hbar|p'|}{2p^{2}}\ll 1 ,
\end{equation}
or
\begin{equation}\label{eq:66}
  \lambdabar = \frac{\hbar}{p} \ll \frac{2(E-V)}{V'(x)},
\end{equation}
where $\lambdabar$ is the de Broglie wavelength divided by $2\pi$. Consequently, the quantization condition of the first-order WKB with half-integer values provides a good approximation.

%%%%%%%%%%%%%%%%%%%%%%%%%%%%%%%%%%%%%%%%%%%%%%%%%%%%%%%%%%%%%%%%%%%%%%

\section{\label{sec:Conclusion} Conclusion} 

We first approximated the wave function as a superposition of two component waves traveling in opposite directions, generating alternately coupled reflection waves under the constraint of Eq. \eqref{eq:9a}. By expanding the amplitude function as a perturbation series, we recovered the Bremmer series. The interactions between the two traveling waves were then incorporated into a sequence of recursive diagonalization steps, yielding a diagonalized a-WKB wave function with an improved momentum (or wave number); in general, this recursive diagonalization mechanism progressively corrects the wave number, leaving only the transmitted part. An accurate approximation requires $n$-fold reflections, for which the $(n+1)$-th term is much smaller than $\mathcal{O}(1)$, so that the resulting expansion forms an asymptotic series that arbitrarily approaches the true wave number as the wave propagates through space. As $\theta$ remains constant along some path, the general quantization formula reduces to the closed form obtained by re-summing the perturbative WKB series to all orders.

Perhaps the most appealing feature of the a-WKB approximation is that it captures information missing from the single WKB 
series, which does not fully account for the reflection effects. The constraint of
\eqref{eq:9a} does not necessitate two highly oscillating waves compared with the vanished amplitude variations over the 
wavelength scale ($a'=b'=0$), but instead 
requires a counterbalance of the overall amplitude variations of two traveling waves ($a'e^{iS}+b'e^{-iS}=0$). 
Such interactions can be geometrically demonstrated in spinor mathematics. However, neither the a-WKB nor the 
diagonalized a-WKB wave function provides a uniform approximation across the classical turning point. This problem 
could potentially be resolved by constructing an associated connection formula or by employing Langer's uniform 
approximation, which expands the wave function in terms of Airy functions; although indirect and highly sophisticated, 
such techniques remain applicable to the classical turning-point problem. {\it Two-component waves} or {\it bipolar waves} provide potentially more appealing resolutions; note that the assumption of a classical momentum in the phase integral 
is insufficiently general, as evidenced by Langer's uniform approximation, which achieves high performance using more 
complicated Airy functions \cite{Berry1972}. As demonstrated in the applications of the 
high-order terms of the a-WKB wave function to the harmonic oscillator problem and to the reflection coefficient of 
the $\text{Sech}(x)^{2}$ barrier, the a-WKB wave function nonetheless embeds the convergence property of the Bremmer series. 
Although we could not identify a physically intuitive method or strategy for passing the turning point, an approximate 
scheme emerged from the development of the a-WKB and diagonalized a-WKB approximations. Adopting the perturbation method, we first developed the a-WKB wave 
function incorporating the Bremmer series to improve the amplitude approximation. We then iterated the coupled differential 
equations and rotated the matrix through a spinor-like transformation, thereby diagonalizing the equation of motion for 
the $\psi_{+}$ and $\psi_{-}$ column vectors and modifying the wave number; in principle, this process can be repeated 
indefinitely. Indeed, our a-WKB approximate scheme resembles the two-component wave formulations, which emphasize approaches 
different from the present study and are scattered throughout the literature on Bohmian mechanics and related areas of 
physics \cite{Bohm1952,Poirier2004reconciling,Floyd1982bohr,Floyd1982modified} (see Appendix \ref{Appendix.C} for a brief 
discussion). For instance, the formulation developed by Floyd shares a motivation similar to ours, namely, solving the 
Schr\"{odinger} equation. He considered a modified potential $U(x)$ smaller than the total energy, $E$, over a finite $x$, 
and formulated an asymptotic relation, $U(x) \to E$ as $x\to \pm \infty$, that resolves the turning-point problem. These 
approaches demonstrate the advantages of the two-component wave formulation, which extracts, from different perspectives, 
the hidden physics in {\it unipolar} wave functions; a brief discussion is given in Sec. C of the Appendix. This study 
highlights the promising possibility of extending the WKB approximation. The correlation among the Floydian formulation, {\it bipolar} wave theory, 
and a-WKB theory, together with solutions to the turning-point problem in our approximate scheme, will be investigated in 
future work.

%%%%%%%%%%%%%%%%%%%%%%%%%%%%%%%%%%%%%%%%%%%%%%%%%%%%%%%%%%%%%%%%%%%%%%%%%%%%%%%%%%%%%%

\begin{acknowledgments}
This research was supported by the National Science and Technology Council of Taiwan, grant number NSTC 112-2221-E-002-141. We thank the National Center for High-performance Computing (NCHC) of Taiwan for providing computational resources, and we thank Prof. Yoshiaki Teranishi for discussions and guidance on classical and quantum mechanics.
\end{acknowledgments}
%%%%%%%%%%%%%%%%%%%%%%%%%%%%%%%%%%%%%%%%%%%%%%%%%%%%%%%%%%%%%%%%%%%%%%%%%%%%%%%%%%%
\appendix

\section{\label{Appendix.A}Coupled Differential Equations of the Amplitudes $a(x)$ and $b(x)$ } 

Assuming the wave function in the ansatz, we obtain
\begin{equation}
  \psi(x) = a(x)e^{iS/\hbar} + b(x)e^{-iS/\hbar},
\end{equation}
where the phase integral is $S(x) = \int^{x}{pdt}$, and $p(x)$ is the classical momentum. Substituting this wave function into the Schr\"{o}dinger equation, we obtain
\begin{equation}
  \begin{aligned}
  &\left(a''+\frac{2ip}{\hbar}a'+\frac{ip'}{\hbar}a\right)e^{iS/\hbar} \\\\
  + &\left(b''-\frac{2ip}{\hbar}a'-\frac{ip'}{\hbar}a\right)e^{-iS/\hbar} = 0.
  \end{aligned}
\end{equation}
Imposing the constraint $ae^{iS/\hbar} + be^{-iS/\hbar}=0$, we obtain
\begin{equation}
  \left(\frac{ip}{\hbar}a'+\frac{ip'}{\hbar}a \right)e^{iS/\hbar} + \left( -\frac{ip}{\hbar}b'-\frac{ip'}{\hbar}b \right)e^{-iS/\hbar}=0.
\end{equation}
We express the derivatives of the amplitudes in terms of each other, specifically writing $a'(x) = -b'(x)e^{-2iS}$ and $b'(x) = -a'(x)e^{2iS}$. We thus obtain two first-order derivatives as coupled differential equations: \\
\begin{equation}
  \left\{
  \begin{array}{rcl}
  &a'(x) &= -\frac{p'}{2p}a(x) + \frac{p'}{2p}b(x) e^{-2iS/\hbar}. \\ \\
  &b'(x) &= -\frac{p'}{2p}b(x) + \frac{p'}{2p}a(x) e^{2iS/\hbar}. 
  \end{array} \right. 
  \end{equation}
which are equivalent to the Schr\"{o}dinger equation under the constraint of Eq. \eqref{eq:9a}. \\\\

%%%%%%%%%%%%%%%%%%%%%%%%%%%%%%%%%%%%%%%%%%%%%%%%%%%%%%%%%%%%%%%%%%%%%%%%%%%%%%%%%%%%%

\section{\label{Appendix.B}Derivation of the Quantization Formula of the Diagonalized a-WKB Wave function}

Applying Keller's approach to the potential well problem, the phase difference condition of the diagonalized a-WKB wave function (Eq. \eqref{eq:44}) is given by
\begin{equation}\label{b1}
  \begin{aligned}
  &\Delta \tilde{S} + \frac{\hbar}{i}\Delta \ln(A) =2K\pi\hbar, \quad (K=0,1,2,\dots).
  \end{aligned}
\end{equation}
Evaluating once around the zeros of the modified momentum along the contour $\Gamma$, the phase difference is obtained as
\begin{equation}\label{b2}
  \oint_{\Gamma}{ \tilde{S}' }dz + \frac{\hbar}{i}\oint_{\Gamma}{ \nabla\ln(A) }dz = 2K\pi\hbar, \quad(K=0,1,2,\dots),
\end{equation}
where the phase and amplitude are respectively given as
\begin{equation}\label{b3}  
  \tilde{S}'(x) = p_{n}(x)     ,
\end{equation}

and

\begin{equation}
  A(x)= \frac{1}{\sqrt{p}}(e^{-\frac{i}{2}\theta \sigma_{y}} e^{ -i\frac{\pi}{4}\sigma_{z} }e^{ \frac{i}{2}\theta_{2}\sigma_{y} }e^{ -i\frac{\pi}{4}\sigma_{z} }\dots)  .
\end{equation}
Note that the distance between the zeros of the amplitude $A(x)$ and the local minimum of the potential (or equilibrium point) shrinks at each diagonalization step. Thus, the effective contour integral always encircles the two zeros of $A(x)$. The phase difference provided by the amplitude is
\begin{equation}\label{b4}
  \begin{aligned}
    &\oint_{\Gamma}{ \nabla \ln (A) }dz \\\\
    &= \oint_{\Gamma}{ \nabla \ln \left[ \frac{1}{\sqrt{p}}(e^{-\frac{i}{2}\theta \sigma_{y}} e^{ -i\frac{\pi}{4}\sigma_{z} }e^{ \frac{i}{2}\theta_{2}\sigma_{y} }\dots) \right] }dz \\\\
    &=\oint_{\Gamma}{ \nabla\ln \left[ ( p^{2}-\frac{\hbar^{2}p'^{2}}{4p^{2}} )( 1- \frac{\hbar^{2}\left|\theta'\right|^{2}}{4\tilde{p}^{2}}  )\dots \right]^{-1/4} \phi(z) dz   }\\\\
    &=-\frac{1}{4}\oint_{\Gamma}{ \nabla \ln\left\{\left[   p^{2}-\frac{\hbar^{2}p'^{2}}{4p^{2}}-\frac{\hbar^{2}\left|\theta'\right|^{2}}{4} + \mathcal{O}(\hbar^{6})+\dots   \right] \right.} \\\\
    &\cdot \phi(z)dz \left. \right\} ,
  \end{aligned}
\end{equation}
\\ where $\phi(x)$ represents the analytical part in the square bracket. The argument in the square bracket can be written as the product of the two simple zeros, $a$ and $b$, of the modified momentum:

  \begin{equation}\label{b5}
    \begin{aligned}
    &\oint_{\Gamma}{ \nabla\ln(A) }dz \\\\
    &= -\frac{1}{4}\oint_{\Gamma}{ \nabla \ln\left\{\left[   p^{2}-\frac{\hbar^{2}p'^{2}}{4p^{2}}-\frac{\hbar^{2}\left|\theta'\right|^{2}}{4}+\dots \right] 
    \phi(z)\right\}dz  } \\\\
    &=-\frac{1}{4}\oint_{\Gamma}{ \nabla \ln[ (z-a)(z-b) ]\tilde{\phi(z)} }dz \\\\
    &=-\frac{1}{4} (2\pi i \times 2) = -\pi i, \quad \text{as $\left| \frac{\hbar p'}{2p^{2}} \right| \ll \mathcal{O}(1)$},
    \end{aligned}
  \end{equation}
\\ where $\tilde{\phi}(x)$ refers to the corresponding analytical function. From the results of Eq. \eqref{b2}, \eqref{b3}, and \eqref{b5}, we have
\begin{equation}\label{b6}
  \frac{1}{\hbar}\oint_{\Gamma}{ p_{n}}dz = (2K+1)\pi, \quad (K=0,1,2\dots).
\end{equation}

\section{\label{Appendix.C} Merits of Two-component Scalar Waves in Quantum Mechanics}

In 1951, Bohm \cite{Bohm1952} applied the following ansatz for the wave function
\begin{equation}\label{eq:c1}
  \psi =R_{B}e^{iS_{B}/\hbar},
\end{equation}
where $R_{B}(x,t)$ and $S_{B}(x,t)$ are real. Applying this ansatz to the time-dependent Schr\"{o}dinger equation yields two governing equations associated with the two fields, $R_{B}(x,t)$and $S_{B}(x,t)$:
\begin{equation}\label{eq:c2}
    \frac{\partial R_{B}}{\partial t} = -\frac{1}{2m}\left[R_{B} \frac{\partial^{2} S_{B} }{\partial x^{2}} + 
    2(\frac{\partial R_{B}}{\partial x})(\frac{\partial S_{B}}{\partial x}) \right]  ,
\end{equation}
and
\begin{equation}\label{eq:c3}
    \frac{\partial S_{B}}{\partial t} = - \left[ \frac{( \partial S_{B}/\partial x )^{2}}{2m}+V(x) - 
    \frac{\hbar^{2}}{2m}\frac{\partial^{2}R_{B}/\partial x^{2}}{R_{B}}  \right] .
  \end{equation}
\\ These two differential equations are nonlinear, and their solutions are correspondingly complicated. As suggested by Bohm, we therefore solved the time-dependent Schr\"{o}dinger equation and inserted the resulting wave function into Eq. \eqref{eq:c1}, thereby obtaining
  \begin{equation}\label{eq:c4}
      \psi = R_{B}\left[ \cos(\frac{S_{B}}{\hbar}) + i \sin(\frac{S_{B}}{\hbar}) \right] ,
  \end{equation}
  and
  \begin{equation}\label{eq:c5}
      S_{B} =  \hbar\tan^{-1}\left[ \frac{\text{Im}(\psi)}{\text{Re}(\psi)} \right].
  \end{equation}
A two-component wave can be obtained by splitting the wave function into its real and imaginary parts. Accordingly, Poirier proposed a more general ansatz, called the {\it bipolar} wave \cite{Poirier2004reconciling}

  \begin{equation}\label{eq:c6}
    \begin{aligned}
      \Psi(x) &= e^{i\delta} r(x)e^{iS_{P}/\hbar} + e^{-i\delta} r(x)e^{-iS_{P}/\hbar} \\
              &\equiv \psi_{+} + \psi_{-},
    \end{aligned}
  \end{equation}
The phase $S_{P}(x)$ and amplitude $r(x)$ can be determined as previously suggested \cite{Poirier2004reconciling}. The linearly independent solutions $\psi_{+}$ and $\psi_{-}$ are complex conjugates with a relative phase of $\delta$, chosen such that the {\it total} wave function, $\Psi(x)$, remains square-integrable. In the stationary state, the quantum mechanical fluxes associated with the two components are equal and opposite, giving a net flux of zero:
  \begin{equation}\label{eq:c7}
    \begin{aligned}
      j_{\pm}(x) &= \frac{\hbar}{2im}\left( \psi_{\pm}^{*}\frac{d\psi_{\pm}}{dx} -\frac{d\psi_{\pm}^{*}}{dx}\psi_{\pm}  \right) \\\\
                 &= \pm \left[ \frac{S'_{P}(x)}{m}r^{2}(x)  \right] \equiv \pm F.
    \end{aligned}
  \end{equation}
Here, $F$ denotes the constant flux of a component. Eq. \eqref{eq:c7} implies the following functional form for the amplitude $r(x)$:
\begin{equation}\label{eq:c8}
  r(x) = \sqrt{ \frac{mF}{S'_{P}} }.
\end{equation}
Therefore, the {\it total} wave function can be written as
\begin{equation} \label{eq:c9}
  \Psi(x) = \sqrt{ \frac{4mF}{S'_{P}} }\cos( \frac{S_{P}}{\hbar} + \delta).
\end{equation}\\
Eq. \eqref{eq:c9} is similar in form to our diagonalized a-WKB wave function, Eq. \eqref{eq:46}, although the two forms differ in their underlying motivation: our scheme attempts to approximate the wave function accurately under appropriate constraints and thereby solve the Schr\"{o}dinger equation, whereas the {\it bipolar} wave aims to resolve practical problems arising from nodes and from large or rapid oscillations of the wave function in ``quantum trajectory methods'' (QTM) \cite{Chou2006computational,Wyatt2005quantum}.

Such formulations, which posit correlations between amplitude and phase, had already been explored by Floyd \cite{Floyd1982bohr,Floyd1982modified} and by Faraggi and Matone
\cite{Faraggi1998equivalence}. The wave function is assumed to be
\begin{equation}\label{eq:c10}
  \psi(x) = \frac{1}{[E-U(x)]^{1/4}}e^{ \pm \frac{i}{\hbar} \int^{x}{\sqrt{2m[E-U(x')]}}dx' }, 
\end{equation}
where the modified potential $U(x)$ satisfies the nonlinear differential equation
\begin{equation}\label{eq:c11}
  U(x) + \frac{\hbar^{2}}{8m}\frac{U''}{E-U} +\frac{5\hbar^{2}}{32m}\left( \frac{U'}{E-U} \right)^{2} = V(x),
\end{equation} 
\\
This equation can be derived by substituting Eq. \eqref{eq:c10} into the time-independent Schr\"{o}dinger equation. The linear combination of Eq. \eqref{eq:c10} therefore provides the {\it total} wave function, under the condition requiring $\Psi(\pm \infty) \to 0$ in the bound state
\begin{equation}\label{eq:c12}
  \Psi(x) = \frac{1}{[E-U(x)]^{1/4}}\sin\left( \frac{1}{\hbar}\int_{\infty}^{x}{\sqrt{2m( E-U )}}dx' \right).
\end{equation}
\\ Under the boundary condition ensuring the asymptotic relation $U(x) \sim E$ as $ x\to \pm \infty $, the turning-point problem is precluded, as indicated by Floyd. The corresponding quantization condition then reduces to an integer-style
\begin{equation}\label{eq:c13}
  \frac{1}{\hbar}\oint{\sqrt{[2mE-U(x')]}dx'} = 2\pi K .
\end{equation}
Although the modified potential $U(x)$ is nonunique, Floyd nonetheless obtained extremely accurate eigenvalue and eigenfunction approximations through numerical calculation for a harmonic oscillator, under suitable initial conditions on $U(0)$ and $U'(0)$.

\nocite{*}

\bibliographystyle{elsarticle-num.bst}
\bibliography{apssamp}% Produces the bibliography via BibTeX.
%\printbibliography

\end{document}